\documentclass{ws-ijmpe}

\begin{document}

\markboth{N. Dinh Dang and N. Quang Hung}{Nuclear pairing at finite
temperature and finite angular momentum}

%%%%%%%%%%%%%%%%%%%%% Publisher's Area please ignore %%%%%%%%%%%%%%%
%\catchline{}{}{}{}{}
%%%%%%%%%%%%%%%%%%%%%%%%%%%%%%%%%%%%%%%%%%%%%%%%%%%%%%%%%%%%%%%%%%%%

\title{NUCLEAR PAIRING AT FINITE TEMPERATURE AND \\
ANGULAR MOMENTUM \footnote{To appear in the Proceedings of the First Workshop on State of the
Art in Nuclear Cluster Physics, Strasbourg 13 - 16 May, 2008.}}

\author{\footnotesize N. DINH DANG }

\address{Heavy-Ion Nuclear Physics Laboratory, Nishina Center for Accelerator-Based Science,\\
RIKEN 2-1 Hirosawa, Wako city, 351-0198 Saitama, Japan \\
and \\ Institute for Nuclear Science and Technique, Hanoi, Vietnam \\
dang@riken.jp}

\author{N. QUANG HUNG \footnote{On leave of absent from
the Institute of Physics and Electronics, Hanoi, Vietnam.} }

\address{Heavy-Ion Nuclear Physics Laboratory, Nishina Center for Accelerator-Based Science,\\
RIKEN 2-1 Hirosawa, Wako city, 351-0198 Saitama, Japan \\
nqhung@riken.jp}

\maketitle

\begin{history}

%\received{(received date)} \revised{(revised date)}
%\accepted{(Day Month Year)}
%\comby{(xxxxxxxxxx)}
\end{history}

\begin{abstract}
An approach is proposed to nuclear pairing at finite
temperature and angular momentum, which includes the effects of
the quasiparticle-number fluctuation and dynamic coupling to
pair vibrations within the self-consistent quasiparticle random-phase
approximation. The numerical calculations of pairing gaps, total
energies, and heat capacities are carried out within a doubly folded
multilevel model as well as several realistic nuclei.
The results obtained show
that, in the region of moderate and strong couplings, the sharp
transition between the superconducting and normal phases is smoothed
out, causing a thermal pairing gap, which does not collapse at a
critical temperature predicted by the conventional
Bardeen-Cooper-Schrieffer's (BCS) theory, but has a tail extended to high
temperatures. The theory also predicts the appearance of a thermally
assisted pairing in hot rotating nuclei.
\end{abstract}

\section{Introduction}

The effect of temperature and angular momentum on pairing
properties is an interesting subject in the study of nuclear
structure. Because of its simplicity, the BCS theory is often used, which
offers a good description of pairing correlation in the
macroscopic systems such as metallic superconductors.
It predicts a collapse
of the pairing gap at $T_{\rm c}$, which signals the sharp superfluid-normal (SN)
phase transition at finite temperature. The BCS theory, however,
ignores quantal and thermal fluctuations, which are significant
in finite small systems. Therefore, it needs to be corrected for the
application to finite nuclei. Various theoretical approaches have been proposed to study the
effects of fluctuations on nuclear pairing~\cite{Moretto2,Goodman,SPA}.
Their results show that, at zero angular momentum, thermal fluctuations smear
out the sharp SN phase transition, resulting in a pairing gap, which does
not collapse at finite temperature. In rotating nuclei, a phenomenon
of temperature induced pair correlations, which reflects the strong
fluctuations of the order parameter in small systems, has also been
predicted~\cite{Frauendorf}. The recent microscopic approach, called
the modified BCS (MBCS) theory~\cite{MBCS} has shown,
for the fist time, that the microscopic source causing the
non-collapsing pairing gap is the quasiparticle-number fluctuation
(QNF).

Recently, we proposed the self-consistent
quasiparticle random-phase approximation (SCQRPA)~\cite{SCQRPA},
which includes the QNF as well as the quantal
fluctuations due to dynamic coupling to pair vibrations.
The purpose of present work is to extend this approach
to finite temperature and finite angular momentum.

\section{Formalism}

The pairing Hamiltonian is considered, which
describes a system of $N$ particles interacting via a pairing
force with the parameter $G$ and rotating with
angular velocity $\gamma$ and a fixed angular
momentum projection $M$ on the laboratory (or body)
fixed $z-$ axis:
\begin{equation}\label{Ha}
     H=\sum_{k}\epsilon_{k}(N_{k}+N_{-k})
-G\sum_{k,k'}P_{k}^{\dagger}P_{k'}-
\lambda \hat{N} -\gamma\hat{M}~,\hspace{5mm} N_{\pm
        k}=a_{\pm k}^{\dagger}a_{\pm k}~,\hspace{5mm}
        P_{k}^{\dagger}=a_{k}^{\dagger}a_{-k}^{\dagger}~,
    \end{equation}
where $a_{\pm k}^{\dagger}$ ($a_{\pm k}$) is the operator that
creates (annihilates) a particle with angular momentum $k$, spin
projection $m_{k}$ or $-m_{k}$, and energy $\epsilon_k$. For simplicity, the subscripts $k$
label the single-particle states $|k,m_{k}\rangle$ with $m_{k}>$ 0,
whereas $-k$  denote the
time-reversal states $|k,-m_{k}\rangle$. The particle
number operator $\hat{N}$ is defined as $\hat{N} =
\sum_k(a_k^{\dagger}a_k + a_{-k}^{\dagger}a_{-k})$, whereas
$\hat{M} =\sum_k m_k(a_k^{\dagger}a_k -
a_{-k}^{\dagger}a_{-k})$ is the $z$-projection of total angular momentum.
The variational procedure is applied to
minimize the expectation value of this Hamiltonian
in the grand canonical ensemble. The result yields
the final equations
for the pairing gap, particle number and total angular momentum,
which include the effect of QNF in the form
    \begin{equation}
       \Delta_{k}=\Delta + \delta\Delta_{k} =
       G\sum_{k'}u_{k'}v_{k'}\langle {\cal D}_{k'}\rangle
       +G\frac{\delta{\cal N}_{k}^2} {\langle {\cal D}_{k}\rangle}u_{k}v_{k}~,
       \label{gapk}
    \end{equation}
    \begin{equation}
       N=2\sum_{k}\bigg[v_{k}^{2}\langle{\cal
       D}_{k}\rangle +\frac{1}{2}\big(1-\langle{\cal
       D}_{k}\rangle\big)\bigg]~, \hspace{5mm}
       M = \sum_k m_k(n_k^{+} - n_k^{-})~,
       \label{NM}
    \end{equation}
    where the quasiparticle energy $E_k$ and renormalized single-particle
    energy $\epsilon_k'$ are given as
        \begin{equation}
       E_{k}=\sqrt{(\epsilon'_{k}-Gv_{k}^{2}
       -\lambda)^{2}+\Delta_{k}^{2}}~,
       \label{Ek}
        \end{equation}
        \begin{equation}
      \epsilon_{k}'=\epsilon_{k}+\frac{G}{\langle{\cal D}_{k}\rangle}
      \sum_{k'}(u_{k'}^{2}-v_{k'}^{2})\bigg(\langle{\cal A}_{k}^{\dagger}{\cal A}_{k'}\rangle
      + \langle{\cal A}_{k}^{\dagger}{\cal
      A}_{k'}^{\dagger}\rangle_{k \neq k'}\bigg)~,
      \label{rspe}
      \end{equation}
with $\langle{\cal D}_k\rangle = 1-n_k^+ - n_k^-$,
and  ${\cal A}^{\dagger}_k \equiv \alpha_k^{\dagger}\alpha_{-k}^{\dagger}$.
The expectation
values $\langle{\cal A}^{\dagger}_k{\cal A}_{k'}\rangle$ and
$\langle{\cal A}^{\dagger}_k{\cal A}^{\dagger}_{k'}\rangle$ are
evaluated by solving a set of coupled equations, which contain the SCQRPA $X$
and $Y$ amplitudes. The QNF is given
as $\delta{\cal N}_{k}^2 = n_k^{+}(1-n_k^{+}) + n_k^{-}(1-n_k^{-})$,
where the quasiparticle
occupation numbers $n_{\pm k}$ are found from the integral equations
    \begin{equation}
        n_{k}^{\pm}=\frac{1}{\pi}\int_{-\infty}^{\infty}\frac{\gamma_{k}^{\pm}(\omega)(e^{\beta\omega}+1)^{-1}}
        {[\omega-E_{k}\pm \gamma m_{k}-M_{k}^{\pm}(\omega)]^{2}+[\gamma_{k}^{\pm}(\omega)]^2}d\omega~,
        \label{nkcoupling}
    \end{equation}
    with the mass operators $M_{k}^{\pm}(\omega)$ obtained by
    solving the set of equations for double-time quasiparticle Green's
    functions and those of a quasiparticle coupled with SCQRPA pair
    vibrations. The quasiparticle dampings are given as
    $\gamma_{k}^{\pm}(\omega)=\Im m[M_{k}^{\pm}(\omega\pm i\varepsilon)]$.

The proposed
approach is called the FTBCS1+SCQRPA theory.  Neglecting the coupling
to SCQRPA, i.e. the factors $\langle{\cal A}^{\dagger}_k{\cal A}_{k'}\rangle$ and
$\langle{\cal A}^{\dagger}_k{\cal A}^{\dagger}_{k'}\rangle$, it
becomes the FTBCS1 theory, which is different from the conventional
FTBCS theory by the presence of the QNF. The violation of particle number at
zero angular momentum is approximately removed by applying the Lipkin-Nogami
(LN) method. The corresponding approaches are called the FTLN1+SCQRPA and FTLN1.

\section{Results}

\begin{figure}
\centerline{\psfig{file=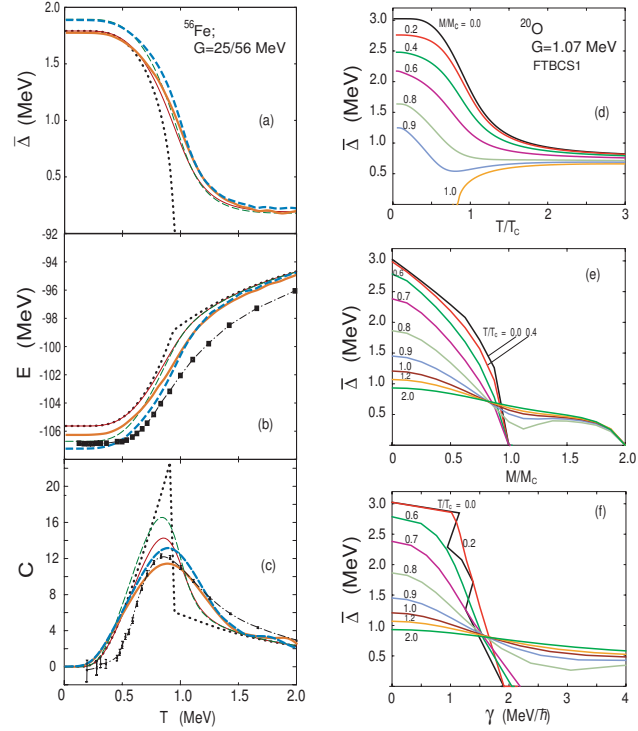,width=8.5cm}} \caption{Left panels:
Pairing gaps (a), total energies (b), and heat capacities (c)
obtained within the FTBCS (dotted lines), FTBCS1 (thin solid lines),
FTLN1 (thin dashed lines), FTBCS1+SCQRPA (thick solid lines) and
FTLN1+SCQRPA (thick dashed lines) for neutrons in $^{56}$Fe. Boxes and crosses
with error bars connected by dash-dotted lines are results of Ref.
$^{8}$.
Right
panels: pairing gaps  as functions of $T$ at different $M$ (d), and as
functions of $M$ (e) and $\gamma$ (f) at various $T$ obtained
within the FTBCS1 theory for neutrons in $^{20}$O. }
\label{fig}
\end{figure}

The numerical calculations are carried out within the $\Omega$
doubly degenerate equidistant model with the number $\Omega$ of levels
equal to that of particles, $N$, as well as for $^{20}$O,
$^{44}$Ca, $^{56}$Fe, and $^{120}$Sn. The results obtained show that,
at zero angular momentum, under the effect of QNF within the FTBCS1
(FTLN1), the sharp SN phase transition predicted by the FTBCS theory
is smoothed out. As the result, the pairing gap does not collapse at
$T = T_{\rm c}$, but has a tail, which extends to high $T$. The
dynamic coupling to the SCQRPA vibrations
significantly improves the agreement with the exact results for
the total energies and heat capacities obtained for $N=10$ as well as
those obtained $^{56}$Fe within the finite-temperature quantum Monte
Carlo method~\cite{Monte} [Figs. \ref{fig} (a) -- \ref{fig} (c)].
However, for heavy nuclei such as $^{120}$Sn, the SCQRPA
corrections are found to be negligible in comparison with the FTBCS1
(FTLN1) results.

For $^{20}$O and $^{44}$Ca, the FTBCS1 pairing gaps, obtained at different
$M$, decreases as $T$ increases and do not collapses at high $T$. At $M$
higher than the critical value $M_{\rm c}$, where the FTBCS gap for
$T = 0$ disappears, there appear thermally
assisted pairing correlations, in which the FTBCS1 gap reappears at a
given $T_1 > 0$, and remains finite at $T > T_1$ [Fig.
\ref{fig} (d)]. This phenomenon is caused
by the QNF within the FTBCS1 theory. At $T = 0$, the QNF is zero, so
the FTBCS and FTBCS1 gaps are the same as functions of $M$ (or $\gamma$),
and both collapse at $M = M_{\rm c}$. However,
with increasing $T$, the FTBCS1 gaps, which are obtained at different $T$, collapse at
$M > M_{\rm c}$, and remain finite even at very high $T$, whereas
those given by the
conventional FTBCS theory vanish at $M \geq M_{\rm c}$ and $T \geq T_{\rm c}$
[Figs. \ref{fig}. (e) and \ref{fig} (f)].

\end{document}